# Oriented 2D Ruddlesden-Popper Metal Halides by Pulsed Laser Deposition

Junia S. Solomon,[1] Nada Mrkyvkova,[2, 2a] Vojtěch Kliner,[1,4] Tatiana Soto-Montero,[1] Ismael Fernandez-Guillen,[3] Martin Ledinský,[4] Pablo P. Boix,[5] Peter Siffalovic,[2] Monica Morales-Masis[1,*]

[1]*MESA+ Institute for Nanotechnology, University of Twente, Enschede, 7500 AE , The Netherlands.
[2]Institute of Physics, Slovak Academy of Sciences, Dúbravská cesta 9, 845 11 Bratislava, Slovakia.
[2a]Center for Advanced Materials Application (CEMEA), Slovak Academy of Sciences, Dúbravská cesta 9, 845 11 Bratislava, Slovakia.
[3]Instituto de Ciencia de los Materiales de la Universidad de Valencia (ICMUV), 46980, Paterna, València, Spain.
[4]Institute of Physics, Academy of Sciences of the Czech Republic, Cukrovarnická 10, Prague , 162 00, Czech Republic.
[5]Instituto de Tecnología Química, Universitat Politècnica València-Consejo Superior de Investigaciones Científicas, Av. dels Tarongers, 46022, València, Spain.

Corresponding author: m.moralesmasis@utwente.nl

Two-dimensional (2D) Ruddlesden-Popper (RP) metal halides present unique and tunable properties. However, direct and oriented synthesis is challenging due to low formation energies that lead to rapid, uncontrolled growth during solution-based processing. Here, we report the solvent-free growth of oriented $n$ = 1 $(PEA)_2PbI_4$ RP films by pulsed laser deposition (PLD). In situ photoluminescence (PL) during deposition reveals formation of the $n$ = 1 phase at the early stages of growth. X-ray diffraction (XRD) and grazing-incidence wide-angle scattering (GIWAXS) confirm a single oriented $n$ = 1 phase, independent of the substrate. Co-localized spatially resolved PL and AFM further validate the conformal growth. While oriented growth is substrate-independent, film stability is not. $(PEA)_2PbI_4$ films grown on strained epitaxial $MAPbI_3$ remain stable for over 184 days without any sign of cation exchange. This work highlights the potential of PLD for direct, room-temperature synthesis of 2D $(PEA)_2PbI_4$ RP films and stable 2D/3D heterostructures.

## Introduction

The search for cutting-edge materials with exceptional optoelectronic properties has sustained enthusiasm in the exploration of three-dimensional (3D) metal halide perovskites (MHPs) after their successful use in photovoltaics. [1–3] Moreover, two-dimensional (2D) layered metal halide perovskites (LMHPs) offer enhanced tunability as a result of the variability in organic spacers and structural complexity. [4–6] These LMHP have been prominently used as surface treatments or as interface/capping layers on top of 3D MHP thin films. [7–9] The primary method for producing these heterostructures is solution processing, which face substantial challenges in managing the distribution of distinct 2D phases due to the minimal difference in their formation energy. [10–12] This issue becomes particularly challenging when growing a 2D layer atop a 3D layer. [13,14] Additionally, solvent compatibility presents a significant problem, as it can damage the underlying layer during the deposition of a 2D/3D layer. [15–20] Lastly, toxic solvents such as N, N-dimethylformamide (DMF) and dimethyl sulfoxide (DMSO) remain a considerable obstacle to green fabrication. [21,22]

While the precise mechanisms governing the formation of these 2D layers are not yet fully understood, vapor-based techniques offer the potential to gain insights and achieve meticulous control over their growth by adjusting various parameters. Additionally, vacuum processing presents a solvent-free alternative for developing multilayer perovskite structures. [23,24] This method eliminates the concern of solvent compatibility and is renowned for depositing high quality semiconductors. [25,26] As a result, vapor deposition techniques have become a focal point of research. Several studies have demonstrated the growth of 2D layers, such as $(BA)_2PbI_4$,[27,28] and $(FPEA)_2PbI_4$,[29] using vapor-assisted processes, where BA represents butylammonium and FPEA stands for fluorinated phenethylammonium. Among these techniques, chemical vapor deposition (CVD) stands out. Dual-source or two-step approaches are often employed to grow 2D layers such as $(PEA)_2PbI_4$, [25,26,30,31] $(PEA)_2PbBr_4$, [26] $(BA)_2PbI_4$ [31] atop 3D MHP structures. However, one-step CVD has typically resulted in the formation of $(BA)_2PbI_4$ and $(PEA)_2SnI_4$ microplates [32] while facing challenges due to the varying sublimation temperatures of precursor powders, which have led to non-conformal layer formations. [31] For the first time, a one-step CVD has successfully achieved growth of all inorganic 2D $Cs_2SnI_2Cl_2$ (Sn-based) and $Cs_2PbI_2Cl_2$ (Pb-based) layers with millimeter-scale coverage. [33] Given the well-established status of vapor phase techniques in the semiconductor industry, investigating 2D layer growth through these methods could facilitate their industrial application. [23] In addition, the creation of conformal and well-defined interfaces in 2D/3D structures requires the implementation of vapor-phase deposition methods, whether in a single or multiple steps. [34]

The direct, one-step growth of 2D LMHPs on top of 3D MHPs by physical vapor deposition methods remains largely unexplored. Our previous research has demonstrated that pulsed laser deposition (PLD), a physical vapor deposition technique, is capable of controlling the growth of epitaxial (epi) thin films of α-$MAPbI_3$ (MA = $CH_3NH_3$).[35] In this study, we leveraged the thickness control and the unique capability of PLD to transfer elements with dissimilar atomic masses/volatilities [36–41] (i.e., mixtures of organic/inorganic components at once) to explore the growth of 2D $(PEA)_2PbI_4$, where PEA denotes phenylethylammonium.

The compound $(PEA)_2PbI_4$ exhibits a Ruddlesden-Popper (RP) structure and is frequently studied for a range of optoelectronic applications. [42] This research investigates its growth by PLD using a dry precursor in a single source. Ex-situ structural analysis of the grown films using specular X-ray diffraction (XRD) and Grazing-Incidence Wide-Angle X-ray Scattering (GIWAXS) allow mapping the formation of the 2D/3D halide perovskite heterostructure. These structural insights are further corroborated by spatially resolved photoluminescence (PL) measurements, which showcase the conformal growth of the 2D $(PEA)_2PbI_4$ on the 3D MHP. Furthermore, atomic force microscopy (AFM) combined with spatially resolved PL is employed to assess the PL distribution of the *n* phases across the 2D $(PEA)_2PbI_4$ layer.

We also investigated the growth of the 2D layer on $SiO_x$/Si substrates and $MAPbI_3$ (011) single crystal (SC) substrates to understand the substrate-induced crystallographic orientation control. The 2D $(PEA)_2PbI_4$ adopts an out-of-plane layered growth regardless of the substrate used. Further, we examined the PL stability of the samples with the underlying 3D MHP, to investigate potential cation exchange behavior. The 2D layer grown on epi-α-$MAPbI_3$ exhibited no formation of higher-order *n* phases, whereas the 2D layer on $MAPbI_3$ SC displayed the formation of higher-order *n* phases over several days. This suggests that epitaxial strain from the underlying epi-α-$MAPbI_3$ may suppress cation migration. Our findings demonstrate the potential of PLD for achieving direct, conformal growth of 2D $(PEA)_2PbI_4$ on 3D MHPs.

# Results

## Direct conformal growth of 2D $(PEA)_2PbI_4$ on 3D $MAPbI_3$ by PLD

The PLD growth of the 2D $(PEA)_2PbI_4$ layer employs parameters optimized in our prior research on the epitaxial growth of α-$MAPbI_3$ using PLD. [35] The alignment of parameters may be due to the appropriate laser fluence (248 nm excimer laser with a fluence of 0.32 $J/cm^2$) for the ablation process involving an organic-inorganic precursor mixture, which is effective even with a large organic spacer cation such as phenethylammonium, ($PEA^+$).

Figure 1.A presents a schematic of the 2D $(PEA)_2PbI_4$ grown atop a epi-α-$MAPbI_3$ (15 nm)/KCl substrate. The growth of the 2D $(PEA)_2PbI_4$ layer is tracked using in situ PL, recording PL spectra in between excimer UV laser pulses, offering real-time insights into electronic structure during the layer formation. Figure 1.B shows the PL spectral feature of the 3D epi-α-$MAPbI_3$ film (Figure S1.A shows the in situ PL) at 750 nm and the 2D perovskite phase ($n$ = 1) at 520 nm. The PL signal from the 2D $(PEA)_2PbI_4$ appears after approximately 300 pulses (i.e. at the initial growth stage) and subsequently increases in time linearly, as shown in the supplementary information (SI), Figure S1.B. Meanwhile, the PL intensity of the 3D MHP increases up to ~1200 pulses, despite its constant thickness, suggesting passivation of the underlying layer. Its subsequent decrease is likely due to increased absorption by the growing 2D layer. While degradation cannot be entirely excluded, we consider it less likely to be the primary cause for the decline in PL intensity of the 3D MHP. [43,44] Only slight variations are observed for the full-width at half-maximum (FWHM) in both the 2D $(PEA)_2PbI_4$ and 3D $MAPbI_3$ cases. Moreover, by increasing the deposition time, we were able to augment the thickness of the 2D layer. Ultimately, the thickness of ~20 nm (1500 pulses), ~40 nm (3000 pulses) and ~70 nm (6000 pulses) of the 2D $(PEA)_2PbI_4$ layer was estimated through cross-sectional scanning electron microscopy (SEM), as shown in Figure S2.

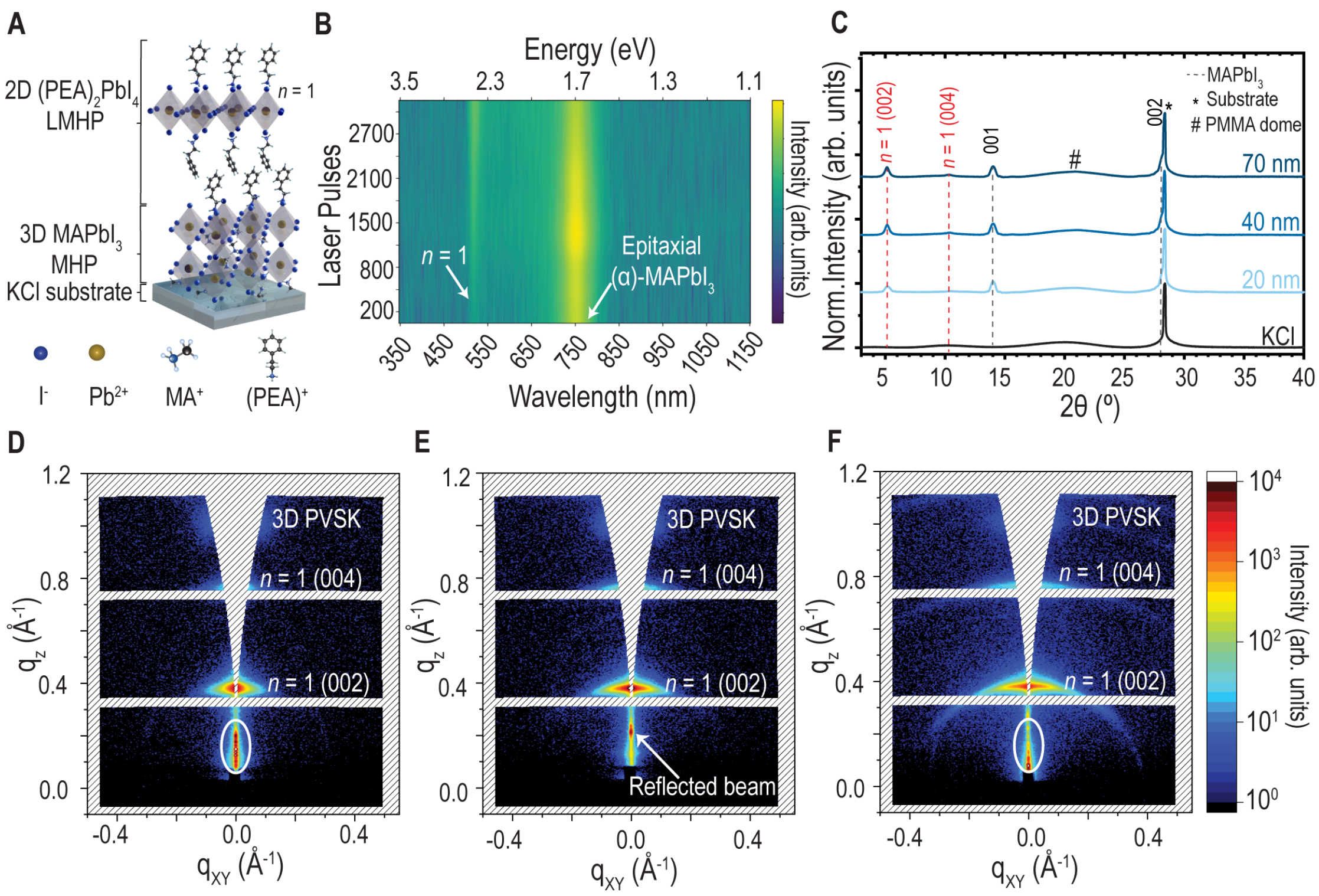

**Figure 1. In-situ photoluminescence during the PLD growth and structural characterization of 2D $(PEA)_2PbI_4$ on epi-α-$MAPbI_3$ by PLD.** (**A**) Schematics of 2D $(PEA)_2PbI_4$ grown on top of 15 nm-thick epitaxial α-$MAPbI_3$ on KCl. (**B**) In situ PL spectra measured during the 2D layer growth. The luminescence of the 2D layer is already visible in the early stages of the growth. (**C**) Specular XRD of 20 nm, 40 nm and 70 nm 2D $(PEA)_2PbI_4$ on epi-α-$MAPbI_3$/KCl. The red dotted line indicates the diffraction peak of the 2D phase and the grey dotted line indicates the peaks from the underlying 3D epi-α-$MAPbI_3$. (**D-F**) GIWAXS patterns of $(PEA)_2PbI_4$ on epi-α-$MAPbI_3$ (15 nm)/KCl with a nominal thickness of (B) 20 nm, (C) 40 nm and (D) 70 nm for the 2D film.

Three different 2D $(PEA)_2PbI_4$ layer thicknesses were deposited to investigate the crystallographic structure and conformality of the 2D $(PEA)_2PbI_4$ layers grown by PLD. For structural analysis, specular X-ray diffraction and GIWAXS were employed. The specular XRD pattern displayed in Figure 1.C indicates the crystal structure of $(PEA)_2PbI_4$, showing a prominent (002) diffraction peak at 5.4°, which is indicative of the $n$ = 1 phase. [26] In addition to the (002) peak, the higher order (004) diffraction peak is faintly visible across all thicknesses of the 2D $(PEA)_2PbI_4$ layer. For a clearer view, a zoomed-in plot is provided in Figure S3 of the SI. Figure 1.D-F shows the GIWAXS diffraction patterns of the respective 2D layers with three different thicknesses. The measured GIWAXS data validate the presence of the 2D $(PEA)_2PbI_4$ $n$ = 1 phase by the (002) and (004) diffractions at 0.38 $Å^{-1}$ and 0.76 $Å^{-1}$, respectively. Furthermore, the diffuse scattering originating from the diffraction peak at $q_z \sim 1$ $Å^{-1}$ reflects the highly (001) textured $MAPbI_3$ perovskite underlayer. We also note that the scattering signal in the region indicated by the white ovals originates from the surface roughness, caused by the terrace-like surface of the KCl substrate. This signal is, however, not related to possible higher-order 2D perovskite phases.

The azimuthally integrated (002) diffraction intensity (see Figure S4.A) scales with 2D perovskite layer thickness, indicating increased 2D phase volume. The spatially confined intensity of the (002) diffraction at $q_{xy} = 0$ $Å^{-1}$ confirms the preferential growth of the 2D $(PEA)_2PbI_4$ RP film with the (001) lattice planes parallel to the substrate. Moreover, the azimuthal distribution of the (002) intensity increases with 2D layers thicknesses (see Figure S4.B-C), indicating the increased mosaicity for the thicker layers.

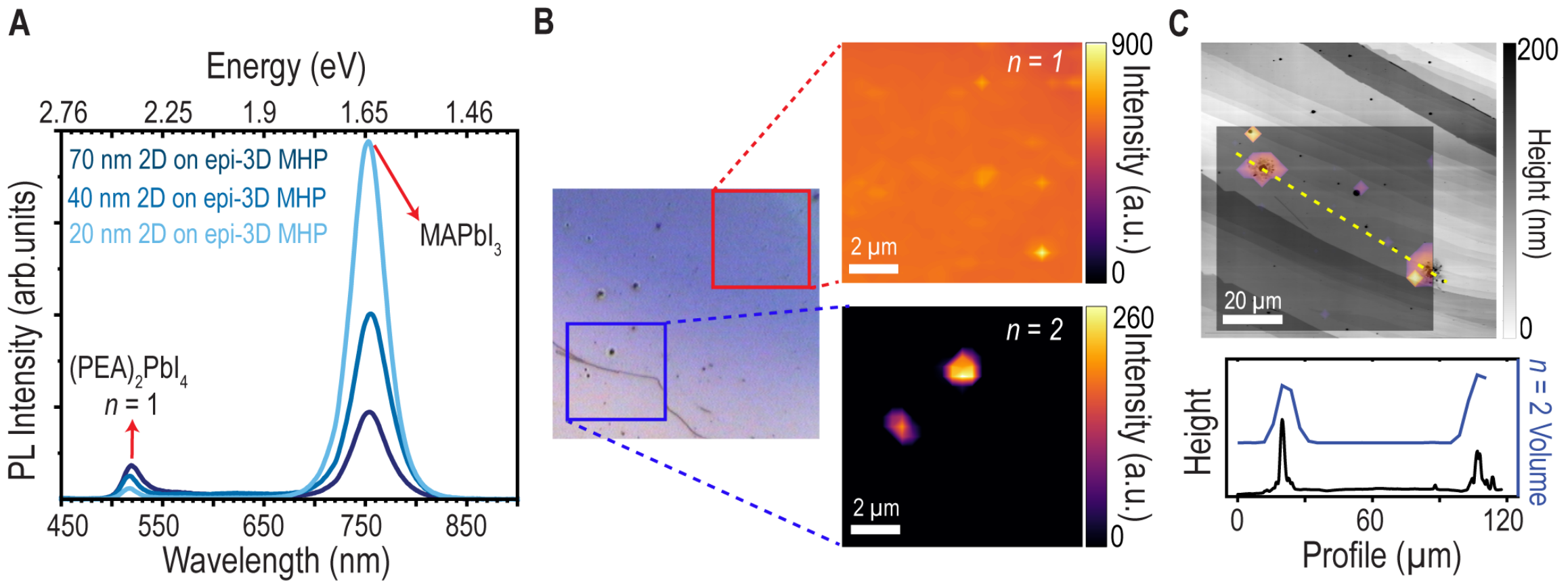

**Figure 2. Conformal coverage of the PLD-grown 2D $(PEA)_2PbI_4$ layer.** (**A**) PL of the 2D $(PEA)_2PbI_4$ layer on epi-α-$MAPbI_3$ (15 nm)/KCl. (**B**) Optical view and the phase volume distributions for $n$ = 1 (red area) and $n$ = 2 (blue area) measured for the 70 nm thick 2D layer. (**C**) Spatial volume distribution of $n$ = 2 phase with a co-localized AFM scan for the 20 nm thick 2D layer. The yellow dashed line depicts a cross-section of the layer's height and the $n$ = 2 phase volume.

Ex-situ PL measurements were performed to analyze the emission properties of the as-grown PLD $(PEA)_2PbI_4$ films. Figure 2.A illustrates a notable PL peak at 520 nm (~ 2.38 eV), indicating the presence of the $n$ = 1 phase. It should be noted that the formation energies of the various low-dimensional phases are quite similar, which has been known to complicate the control over phase formation with other techniques as reported in ref.[45]. Notably, for the PLD grown films the $n$ = 1 phase dominates in the PL spectra.

Since the 2D $(PEA)_2PbI_4$ layer is grown directly on the 3D MHP layer, it is crucial to confirm its conformality, i.e., the 2D phase volume and its spatial distribution. [46] Although XRD, GIWAXS, and PL measurements reveal the presence of the 2D $(PEA)_2PbI_4$ film, it might not necessarily be a conformal layer. To gain further insight into the conformality and emission intensity of the $(PEA)_2PbI_4$ film, spatially-resolved PL was carried out, along with co-localized AFM (Figure 2.B-C). Figure 2.B shows the spatially-resolved PL maps for the 70 nm thick $(PEA)_2PbI_4$ layer, where the greatest amount of the 2D perovskite phase is expected. The PL scans were performed in two distinct areas, while PL emission spectra were measured at each point, and the PL peaks were fitted by a Gaussian function (for more details on the fitting, Figure S5). The PL peak area then represents the volume of the corresponding 2D perovskite phase. In the red area (see Figure 2.B), with no pronounced particulates (circular dark spots, that are generated during the PLD process. [47,48]) visible in the optical microscope image, only the $n$ = 1 phase was detected, forming a continuous layer with a relatively homogenous distribution. On the other hand, in the blue area, where a few particulates are visible, the $n$ = 2 phase was observed in the PL spectra (peak at ~ 568 nm), along with $n$ = 1. Interestingly, the $n$ = 2 perovskite phase is predominantly where the particulates were located. We note that the $n$ = 2 phase was also observed for thinner $(PEA)_2PbI_4$ layers (see Figure S6), with the same character of spatial distribution at the particulates.

To further investigate the topographical character of the particulates where the higher-order 2D perovskites are located, we performed AFM imaging colocalized with the PL scans. Figure 2.C shows the overlay of an AFM scan and a PL map for $n$ = 2 phase. Clearly, the $n$ = 2 phase is not distributed homogenously at the surface and is located solely at particulates.. Moreover, the GIWAXS pattern (Figure 1.D-F) shows no evidence of the higher 2D perovskite phases, not even for the 70 nm thick layer. This confirms that the higher $n$-phase volume is marginal.

## PLD growth of 2D $(PEA)_2PbI_4$ RP films on different substrates

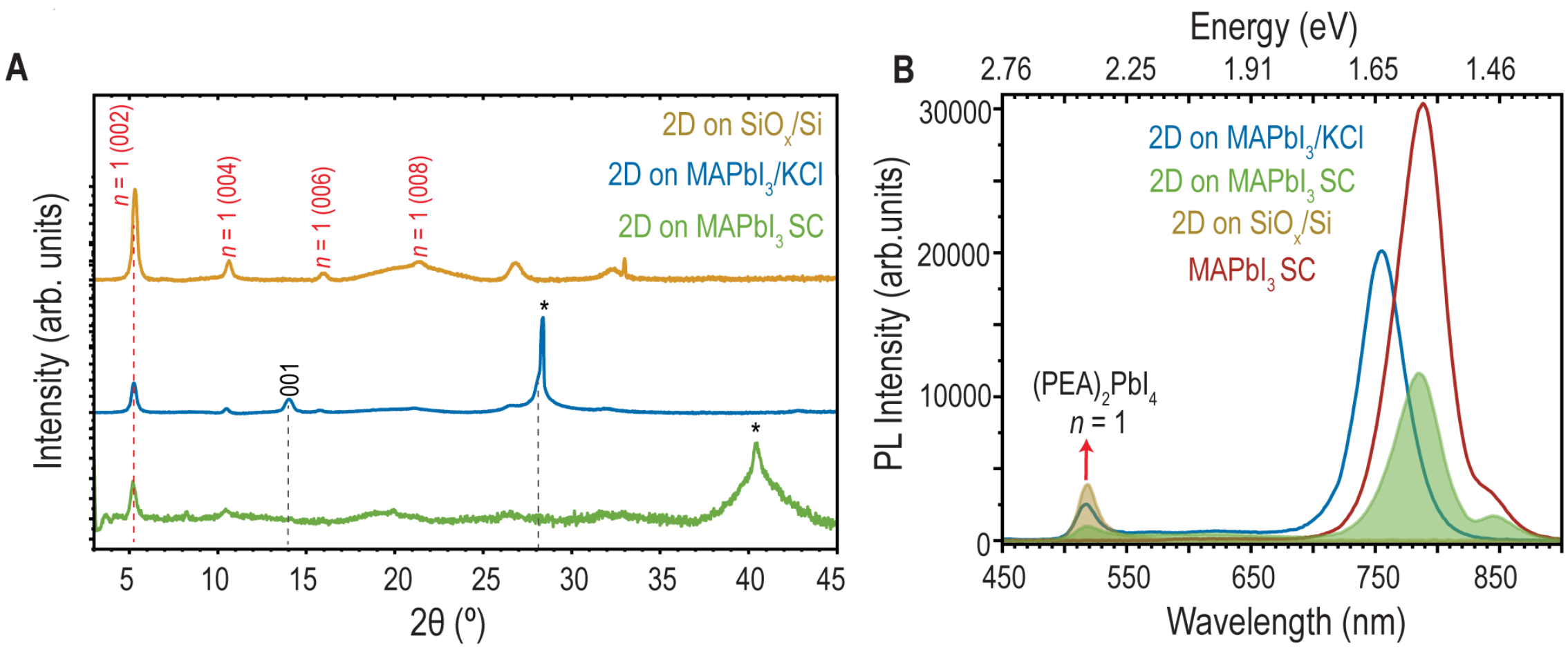

**Figure 3. PLD-grown 2D $(PEA)_2PbI_4$ layer on various substrates.** (**A** and **B**) show the specular XRD and the PL spectra of 40 nm thick 2D $(PEA)_2PbI_4$ layer grown on $SiO_x$/Si, epi-α-$MAPbI_3$/KCl and $MAPbI_3$ SC, respectively. In Figure 4 (A), the red dotted line indicates the (002) diffraction peak of the 2D $(PEA)_2PbI_4$ phase. The ∗ indicate the diffraction peak of the $MAPbI_3$ SC substrate (green line) and the KCl substrate (blue line) and the grey dotted line indicates the peaks from the underlying epi-α-$MAPbI_3$. The 2D $(PEA)_2PbI_4$ layer grown on $MAPbI_3$ SC was evaluated using an alternative XRD setup, as the thickness of the $MAPbI_3$ SC (~3 mm) made it unsuitable for analysis in Bruker D8; for details, refer to the methods section.

Having demonstrated the growth of the 2D layer by PLD, we have further examined the growth of the 2D $(PEA)_2PbI_4$ layers on various substrates to determine the impact of the substrate on the film growth. A non-crystalline Si with native $SiO_x$ wafer and $MAPbI_3$ (011) single crystal (SC) were chosen in addition to the epi-α-$MAPbI_3$(001) (15 nnm)/KCl. It was anticipated that altering the surface energy and/or crystallographic orientation of the substrate would affect the 2D growth. However, the XRD patterns of the 2D $(PEA)_2PbI_4$ layer shown in Figure 3.A, indicate that the 2D $(PEA)_2PbI_4$ layer retains a (001) texture regardless of the substrate employed. The XRD patterns and the rocking curve of the bare crystalline substrates (KCl and $MAPbI_3$ SC) are shown in Figure S8.A and B, respectively. Further, GIWAXS data of the 2D layer grown on the $MAPbI_3$ (011) SC shows the presence of the $n$ = 1 phase (Figure S9.A).

The PL for these samples further confirms the formation of the $n$ = 1 2D phase (Figure 3.B). The shoulder observed for the $MAPbI_3$ SC, in addition to the main emission of $MAPbI_3$ at 1.56 eV, arises as a result of self-absorption.[49] Based on the XRD pattern, the preferred (001) growth orientation of $(PEA)_2PbI_4$, regardless of the substrate, can be attributed to the intrinsic energetic favourability and structural stability of the (001) orientation of $(PEA)_2PbI_4$, as commonly observed for layered MHPs. This suggests a van der Waals like interaction at the interface, rather than strong chemical bonding with the substrate, governing the growth process. This allows the 2D layer to adopt its preferred orientation independent of the underlying substrate.[50–52] Additionally, the growth orientation of the $(PEA)_2PbI_4$ films remains unaltered, i.e. (001) textured, even when grown at varying deposition pressures, as demonstrated in Figure S10.

## Stability monitoring of the 2D $(PEA)_2PbI_4$ layer

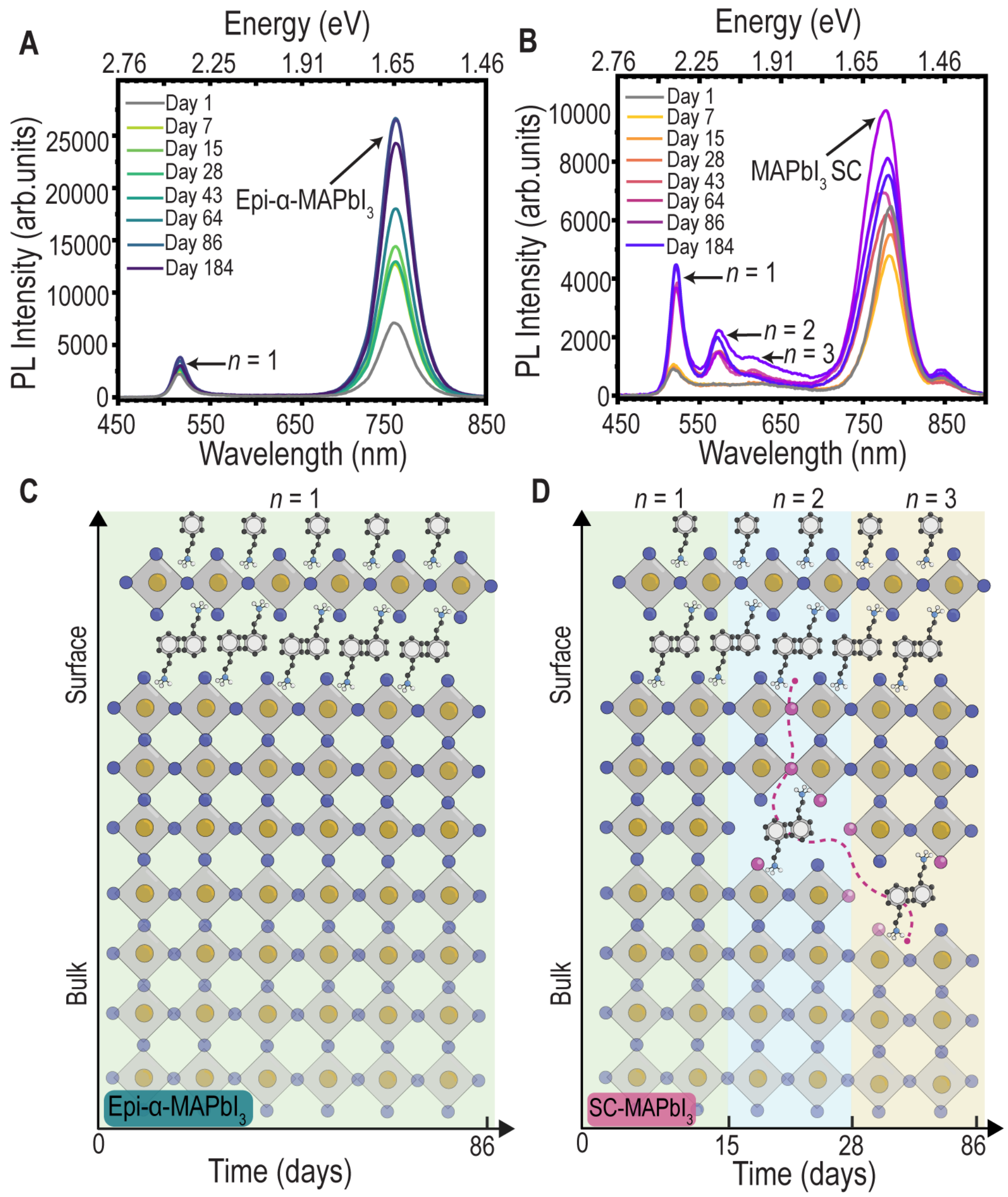

**Figure 4. Photoluminescence stability of 2D $(PEA)_2PbI_4$ deposited on various substrates measured over time**. (**A**) PL spectra of 2D $(PEA)_2PbI_4$ (40 nm)/epi-α-$MAPbI_3$ (001) (15 nm)/KCl and (**B**) 2D $(PEA)_2PbI_4$/$MAPbI_3$ (011) SC .The samples were stored and measured inside the glovebox without it being exposed to ambient conditions. (**C-D**) Schematics of cation migration in the two samples, 2D $(PEA)_2PbI_4$ (40 nm)/epi-α-$MAPbI_3$ (001) (15 nm)/KCl and 2D $(PEA)_2PbI_4$/$MAPbI_3$ (011) SC, respectively.

Once the $n = 1$ phase and preferential orientation of the 2D $(PEA)_2PbI_4$ layer on several substrates were confirmed, we proceeded to examine the stability of the deposited $(PEA)_2PbI_4$ (nominal thickness of 40 nm), with a particular focus on the cases of the epi-α-$MAPbI_3$ and $MAPbI_3$ SC substrates. The aim was to assess the $n = 1$ phase stability via PL measurements, as the presence of the underlying 3D MHP could potentially induce ion migration or MA-PEA cation exchange behavior. [53–56] To investigate this, we monitored the PL over time while keeping the samples in a $N_2$ glovebox to eliminate moisture-related degradation and isolate the effects of cation migration.

Figure 4.A and B shows the PL spectra over time for the 2D $(PEA)_2PbI_4$ layers deposited on the epi-α-$MAPbI_3$ and $MAPbI_3$ SC, respectively. Initially, both samples exhibit the $n = 1$ phase. However, over time, the higher-order $n$ phases emerge only in the $MAPbI_3$ SC sample, while the epi-α-$MAPbI_3$ sample remains stable. The emergence of these higher-order phases in $MAPbI_3$ SC coincides with the formation of a thin-film-like $MAPbI_3$ layer, as indicated by the appearance of a new photoluminescence band at 755 nm (Figure S11. B). This layer starts forming at day 7 and progressively thickens over time (Figure S11.A and B), suggesting that the $(PEA^+)$ cation migrates from the 2D $(PEA)_2PbI_4$ layer into the $MAPbI_3$ SC. This observation aligns with the formation of higher-order $n$ phases for the 2D on $MAPbI_3$ SC. This interpretation is further supported by microscopic studies (Figure 2.B-C) that revealed the localized formation of higher-order 2D phases solely at particulate sites. These sites likely correspond to small bulk $MAPbI_3$ particles remaining from the deposition process.

The difference in stability between the two substrates suggests that strain plays a critical role in regulating cation migration. Previous studies have shown that strain can supress ion migration by increasing the formation energies of halide vacancies, with the highest suppression occurring under compressive strain. [57,58] In the case of epi-α-$MAPbI_3$, the epitaxial strain likely raises the energy barrier for cation migration, preventing $PEA^+$ diffusion and stabilizing the $n = 1$ phase. In contrast, $MAPbI_3$ SC, grown by the inverse solubility method, [59] lacks such strain constraints, allowing cation migration to proceed more readily. This interpretation aligns with studies demonstrating the critical influence of strain on ion migration in metal halide perovskites. [60,61]

To confirm that $PEA^+$ migration originates only from the top facet rather than the side facets of $MAPbI_3$ SC (during deposition), we performed PL measurements from the backside of a fresh sample and compared it to a similar sample after 86 days (Figure S11.C). The results indicate that the $PEA^+$ diffuses from the top surface, forming higher-order phases due to ion migration. The absence of such phase evolution in the epi-α-$MAPbI_3$ sample reinforces the hypothesis that epitaxial strain supresses cation migration, thereby stabilizing the 2D layer over time. To further support our hypothesis, we also monitored the stability of the 2D $(PEA)_2PbI_4$ film grown on a $SiO_x$/Si substrate, which lacks an underlying 3D MHP layer. As expected, no notable PL evolution or phase change was observed over time (SI, Figure S12).

# Discussion

We report the successful direct dry synthesis of 2D $(PEA)_2PbI_4$ RP films via PLD. Our findings highlight PLD's ability to precisely control the growth of 2D/3D perovskite heterostructures, enabling fine-tuning of both film thickness and interface quality. Structural characterization by XRD and GIWAXS confirmed the presence of the $n = 1$ phase of the 2D $(PEA)_2PbI_4$ films and its preferential texture. Additionally, in situ PL measurements during PLD growth indicated that the 2D layer begins crystallizing from the initial stages of the deposition. Spatially resolved PL co-localized with AFM topography further confirmed that the 2D $(PEA)_2PbI_4$ layer forms a conformal layer over the 3D perovskite. When grown on other substrates such as $SiO_x$/Si and $MAPbI_3$ single crystal (011), the 2D $(PEA)_2PbI_4$ preserves its (001) preferential orientation, independent of the substrate. When deposited on epitaxially strained α-$MAPbI_3$, the 2D $(PEA)_2PbI_4$ layer maintains its $n = 1$ phase, showing a stable PL intensity over time. In contrast, deposition of the 2D $(PEA)_2PbI_4$ layer on strain-free $MAPbI_3$ single crystal results in gradual formation of higher-order $n$ phases indicating cation migration. This highlights the role of epitaxial strained α-$MAPbI_3$ in suppressing cation migration or exchange between the 2D and 3D phases. In conclusion, this study demonstrates the potential of PLD as a solvent-free method for fabricating high-quality, direct synthesis of conformal 2D layers on top of 3D perovskites. Our findings pave the way for further exploration of vapor-phase methods for developing 2D/3D heterostructures, unlocking new opportunities for stable, high-performance optoelectronic materials.

**Methods**

**Pulsed Laser Deposition (PLD)**
PLD was executed using a Coherent KrF excimer laser ($\lambda$= 248 nm) to ablate the solid target within a customized (TSST Demcon) vacuum chamber under an argon atmosphere with a working pressure of 0.03 mbar. Repeated depositions were carried out on a newly ground area of the PLD target (described below). The target- to-substrate distance was maintained at 55 mm for all depositions. All procedures were conducted at room temperature (approximately 25 °C). The laser fluence was consistently set at 0.32 J/cm$^2$ for each sample unless noted otherwise, with a frequency of 1 Hz for the laser pulses: 6000, 3000, and 1500 pulses, resulting in film (nominal) thicknesses of 70 nm, 40 nm, and 20 nm, respectively. The laser spot size was 1.24 $mm^2$.

**PLD Targets**

Phenethylammonium iodide (PEAI, CAS-No:151059-43-7, 98% purity, purchased from Greatcell Solar Materials) and lead iodide ($PbI_2$, 99.999% purity, purchased from Sigma-Aldrich) powders were weighed using an analytical balance (±0.01 mg) inside an $N_2$-filled glovebox. A non-stoichiometric ratio of the inorganic to the organic components ($PbI_2$:PEAI = 1:8) was used. The powders were then transferred to zirconia-coated vials and mixed for 48 hours using a homemade rotatory ball mill with zirconia balls. Subsequently, the mixed powders were uniaxially pressed at room temperature into a ~2.5-mm-thick disc with a diameter of 20 mm. The PLD targets were pre-ablated in situ before each deposition and polished in between depositions in an $N_2$ glovebox.

**Substrates**

Three type of substrates were used for the growth of $(PEA)_2PbI_4$ by PLD, namely, epitaxial $MAPbI_3$ on KCl, Si wafers with native oxide and single crystal $MAPbI_3$ substrates. The preparation of the epitaxial $MAPbI_3$ on KCl is described in detail in ref [35], and others are described below:
Si substrates: cleaned prior deposition by ultrasonic cleaning in an acetone bath, followed by a rinse with isopropanol and then dried using $N_2$ for 10 minutes each step.
$MAPbI_3$ single crystal: Lead(II) iodine ($PbI_2$, purity of 99.99%, acquired from Tokio Chemical Industri, TCI), methylammonium iodine (MAI, purity of 99.99%, purchased from Greatcell Solar Materials),and γ-Butyrolactone (GBL, purity of 99%, obtained from Tokio Chemical Industri, TCI) were utilized as the precursor materials. Lead (II) iodine and methylammonium iodine were dissolved in GBL at a concentration of 1 M. The mixture was kept stirring at 50 ºC until the precursors were fully dissolved. The solution was then filtered through a 0.2 mm pore-size filter. After filtration, the resulting solution was placed into a sealed 15 $cm^3$ vial and immersed in an oil bath preheated at 50 ºC. The temperature was gradually increased at a rate of 10 °C/h until it reached 110 °C. Finally, the solution was maintained at 110 °C for a period of 24 h. This approach resulted in crystals with uniform size.

**X-ray Diffraction (XRD)**

Measurements were carried out in a symmetric configurations using a Bruker D8 Discover X-ray diffractometer. The instrument is equipped with a rotating anode turbo X-ray source (TXS) combined with a hybrid parallel focusing mirror (Montel) and a two-bounce channel-cut germanium monochromator to further reduce the divergence of the incident beam. A circular beam collimator of 1 mm was used for out-plane XRD measurements. The step size for all the measurements was 0.01°. However, the time per step for all measurements are different. Hence, the intensity values are not absolute and are reported in arbitary units (arb.units).
The samples grown on the $MAPbI_3$ SC were measured in a symmetric configuration using a PANalytical X'Pert PRO with a Cu anode X-ray source.

**Grazing-Incidence wide-angle X-ray scattering (GIWAXS)**

2D reciprocal space maps were measured utilizing a custom-designed WAXS/SAXS system equipped with a MetalJet X-ray source (Excillum) with an X-ray energy of 9.25 keV (λ = 1.34 Å) and the Pilatus 300 K detector (Dectris). Samples were measured in vacuum, with an angle of incidence of 1.5°. The acquisition time was set to 60 seconds.

**Photoluminescence (PL)**

Measurements were performed in air ( 35% RH) using a home-built PL setup consisting of a 405 nm laser diode module, 152 mW (Matchbox series from Integrated Optics), and a StellarNetBLUE-Wave Spectrometer coupled with a fiber optic.

In-situ PL spectra were measured with a long-distance custom-built PL setup designed using the Thorlabs cage system and Thorlabs optical components. As an excitation source, a focused 405 nm collimated laser module with a constant power of 4.5 mW (elliptical beam) was employed. The distance between a sample and the objective lens was set to 301 mm. A fiber-coupled StellarNet BLUE-Wave Spectrometer was used.

**Spatially resolved Photoluminescence and colocalized PL-AFM imaging**
The PL scans were measured on a confocal microscope (Alpha 300R, WiTec, Germany), using a 405 nm excitation laser wavelength. The laser intensity was set to 60 nW so as not to cause the degradation of perovskite 2D phases. The PL signal was collected by a 50× objective and detected by a Peltier-cooled CCD camera. The scanned area was 100 × 100 $\mu m^2$, and the PL spectrum was measured with a step of 2 μm. The PL maps were generated for $n = 1$ and $n = 2$ phases by fitting the measured spectral maps with a Gaussian peak function. The colocalized AFM scans were measured in a tapping mode, using a silicon tip (RTESPA-300, Bruker). The measurements were performed in air.

**Declaration statements:**

**Availability of data and materials**

All the data needed to evaluate the conclusions of the paper are presented in the paper and/or the Supplementary Materials. Additional data related to this paper may be requested from the authors.

**Acknowledgements**
J.S.S, T.S.M. and M.M.M acknowledge the research funding support from the European Research Council (ERC) under the European Union's Horizon 2020 Research and Innovation Program (CREATE, Grant Agreement No. 852722) received by M.M.M. We authors acknowledge the technical support of TSST Demcon, Dominic Post, and Jeroen S. van Valkenhoef for their technical assistance and Daniel Monteiro Cunha for providing the code to visualize the in-situ PL measurements. We thank Mark Smithers and Melissa Goodwin for the SEM and FIB-SEM measurements. N.M is grateful for the support provided by the IMPULZ program of the Slovak Academy of Sciences, project no. IM-2023-82, Slovak Research and Development Agency (APVV-21-0297, SK-CZ-RD-21-0043), and JRP/2023/727/PVKSC. V.K and M.L. would like to acknowledge the projects GACR 24-11652S and PVKSC 9F23003, as well as the use of the CzechNanoLab research infrastructure supported by the MEYS (LM2023051). P.P.B thanks Generalitat Valenciana for the funding via Pla Gent-T (grant ESGENT 010/2024).

**Author contributions**
JSS and MMM conceived the idea. JSS prepared the thin films by PLD and performed the XRD, steady-state PL measurements, and analysis. JSS and VK did the in situ PL measurements. NM and PS performed the GIWAXS, spatially resolved PL and AFM scans. IFG and PPB synthesized the $MAPbI_3$ single crystals. JSS and MMM performed the investigation. JSS, NM, TSM and MMM worked on the visualization of the results. NM, PS, ML and MMM supervised the overall work. JSS and MMM wrote the manuscript with the input from all the authors. All coauthors analyzed and discussed the results.

**Competing interests**
The authors declare that they have no competing interests.

**Ethics and inclusion**
We support the inclusive, diverse and equitable conduct of research.

**Supplementary information**
The article has an accompanying supplementary file.

# Oriented 2D Ruddlesden-Popper Metal Halides by Pulsed Laser Deposition

**This PDF includes:**

## Direct conformal growth of 2D $(PEA)_2PbI_4$ on 3D $MAPbI_3$ by PLD

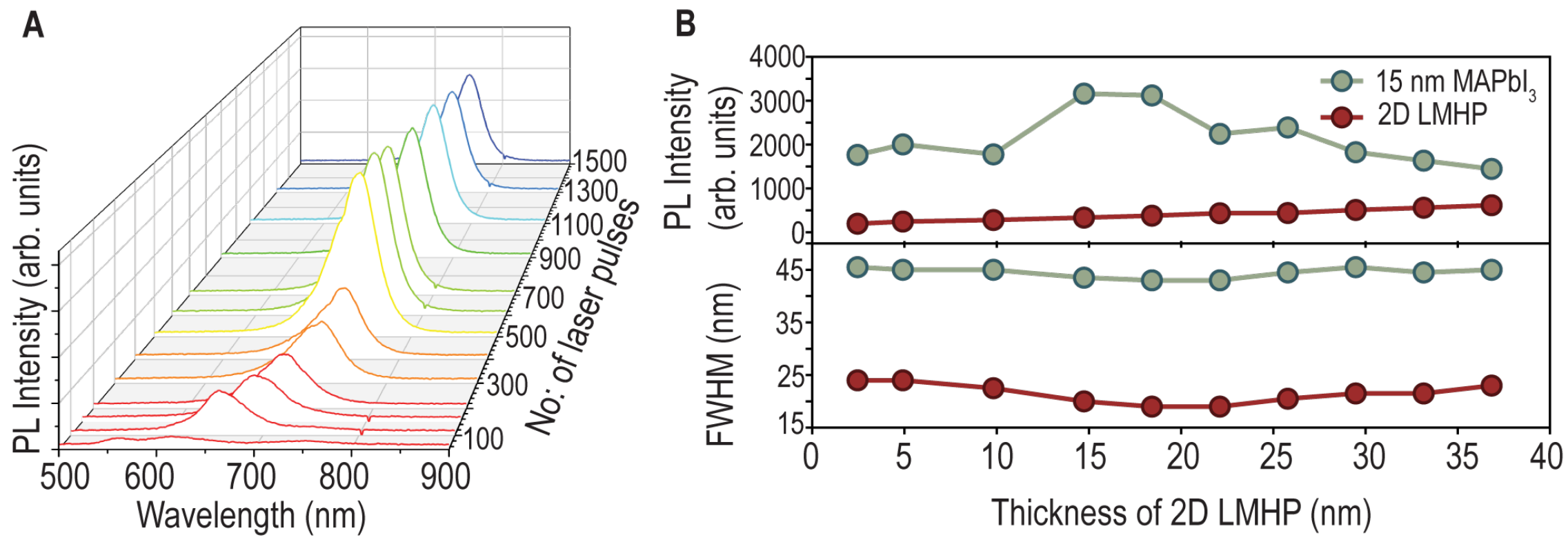

**Figure S1: In situ PL of 3D epi-α-$MAPbI_3$ and 2D $(PEA)_2PbI_4$** (**A**) In situ PL performed during the growth of the underlying epitaxial α-epi-$MAPbI_3$ on KCl. (**B**) PL intensity and full-width-half-maxima (FWHM) of the 15 nm $MAPbI_3$ (750 nm) and the 2D $(PEA)_2PbI_4$ (520 nm) measured across various thicknesses of the 2D layer. The thickness values were interpolated based on cross-section SEM, shown in Figure S2.

In figure S1(A), the blue shift of the $MAPbI_3$ peak can be associated with the quantum size effect during the initial stages of the growth. [1]

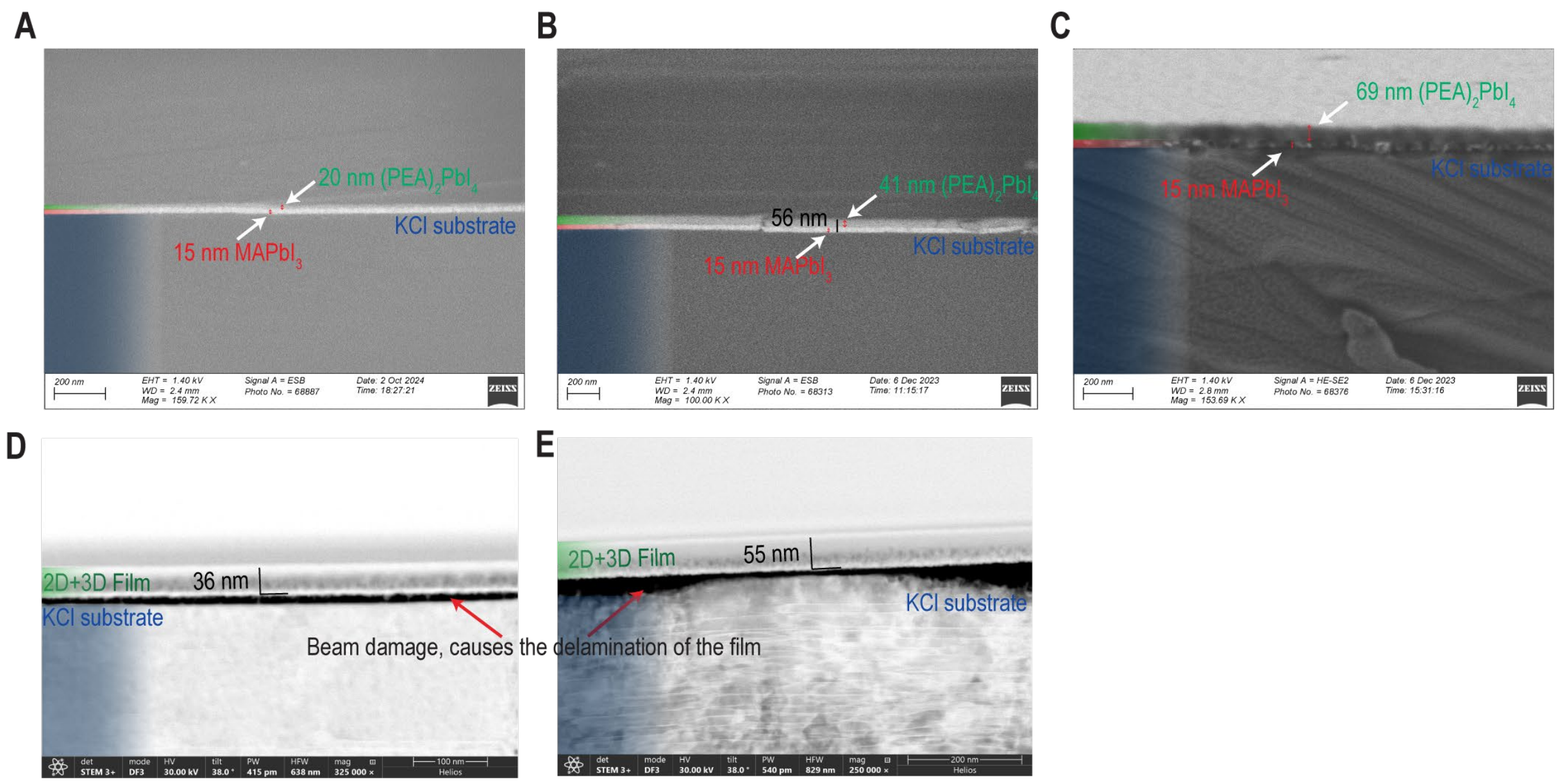

**Figure S2: Thickness estimation from cross-section SEM of the 2D $(PEA)_2PbI_4$ on top of the 3D epi-α-$MAPbI_3$/KCl,** (**A**) for the 2D layer grown with 1500 laser pulses, (**B**) for the 2D layer grown with 3000 laser pulses, and (**C**) for the 2D layer grown with 6000 laser pulses. (**D**) FIB-SEM of for the 2D layer grown with 1500 laser pulses, and (**E**) for the 2D layer grown with 3000 laser pulses.

Accurate measurement of the2D $(PEA)_2PbI_4$ layer thickness, particularly for the thinner films, was challenging due to charging effects caused by the insulating KCl substrate. High-resolution imaging using the HE-SE2 detector was significantly affected by charging effects. To address this, we utilized the ESB (energy-selective backscatter) detector mode, which provides enhanced compositional contrast and is more tolerant to charging. We also attempted cross-sectional FIB-SEM imaging to obtain higher magnification and clarity. However, due to the beam sensitivity, the film delaminates from the KCl substrate, as shown in Figure S2.D-E. Therefore, the most reliable approach for estimating the 2D layer thickness was to measure the total thickness of the combined 2D + 3D stack and subtract the known thickness of the underlying 3D $MAPbI_3$ layer. This methos yielded estimated 2D thicknesses of approximately 20 nm, 40nm, and 70 nm, for 2D layer grown with 1500, 3000, and 6000 laser pulses.

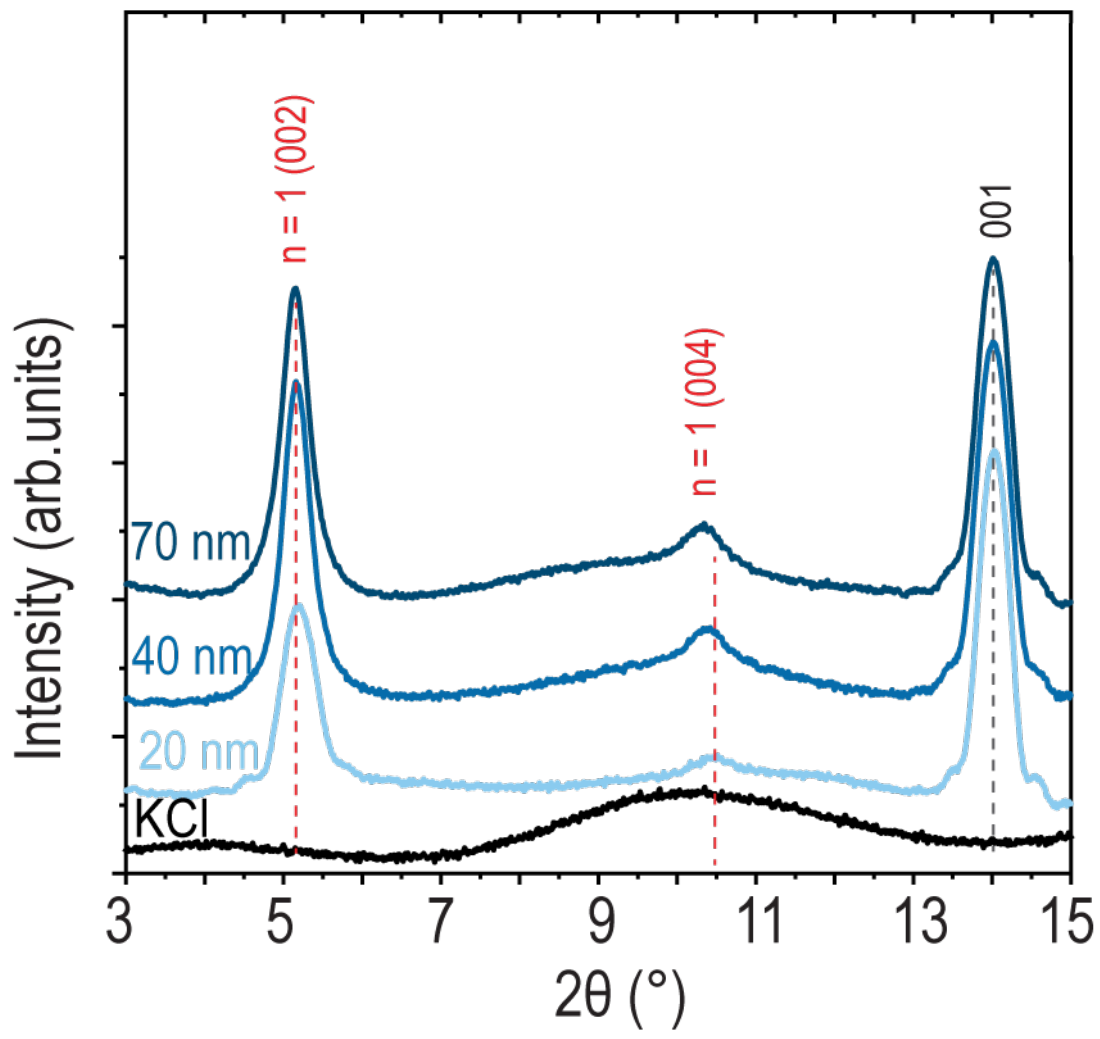

**Figure S3: Specular XRD of 20 nm, 40 nm and 70 nm 2D $(PEA)_2PbI_4$ on epi-α-$MAPbI_3$/KCl.** It is a zoom-in of the Figure 1.C with 2θ range from 3° to 15°. The red dotted line indicates the diffraction peak of the 2D phase and the grey dotted line indicates the peaks from the underlying 3D epi-α-$MAPbI_3$.

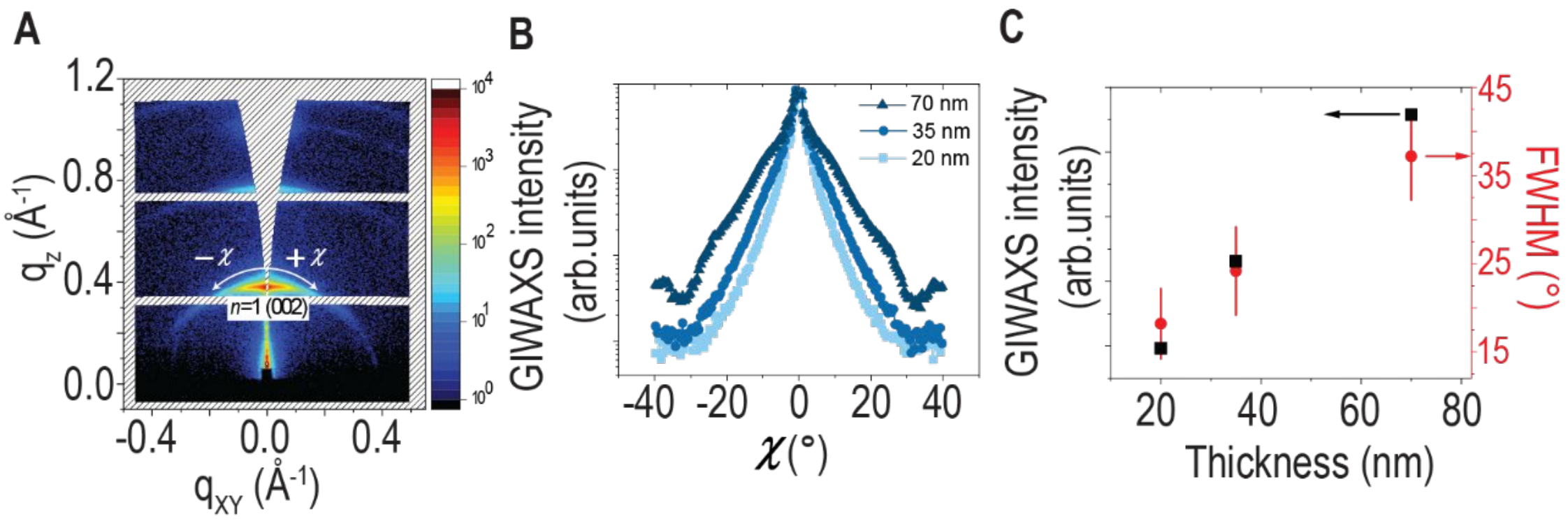

**Figure S4: GIWAXS structural characterization of the $(PEA)_2PbI_4$ layers on epi-α-$MAPbI_3$/KCl by PLD.** (**A**) GIWAXS patterns of 70 nm thick $(PEA)_2PbI_4$ layer on α-epi-$MAPbI_3$ (15 nm)/KCl substrate. The white arrows indicate the direction of the azimuthal angle $\chi$. (**B**) Azimuthal distribution of the (002) diffraction intensity (in the logarithmic scale) for different thicknesses of $(PEA)_2PbI_4$ layers indicating the increased mosaicity with increasing layer thickness. (**C**) 2D perovskite thickness dependence of the integrated (002) diffraction intensity and the full width at half maximum (FWHM) obtained by fitting of the curves shown in (B).

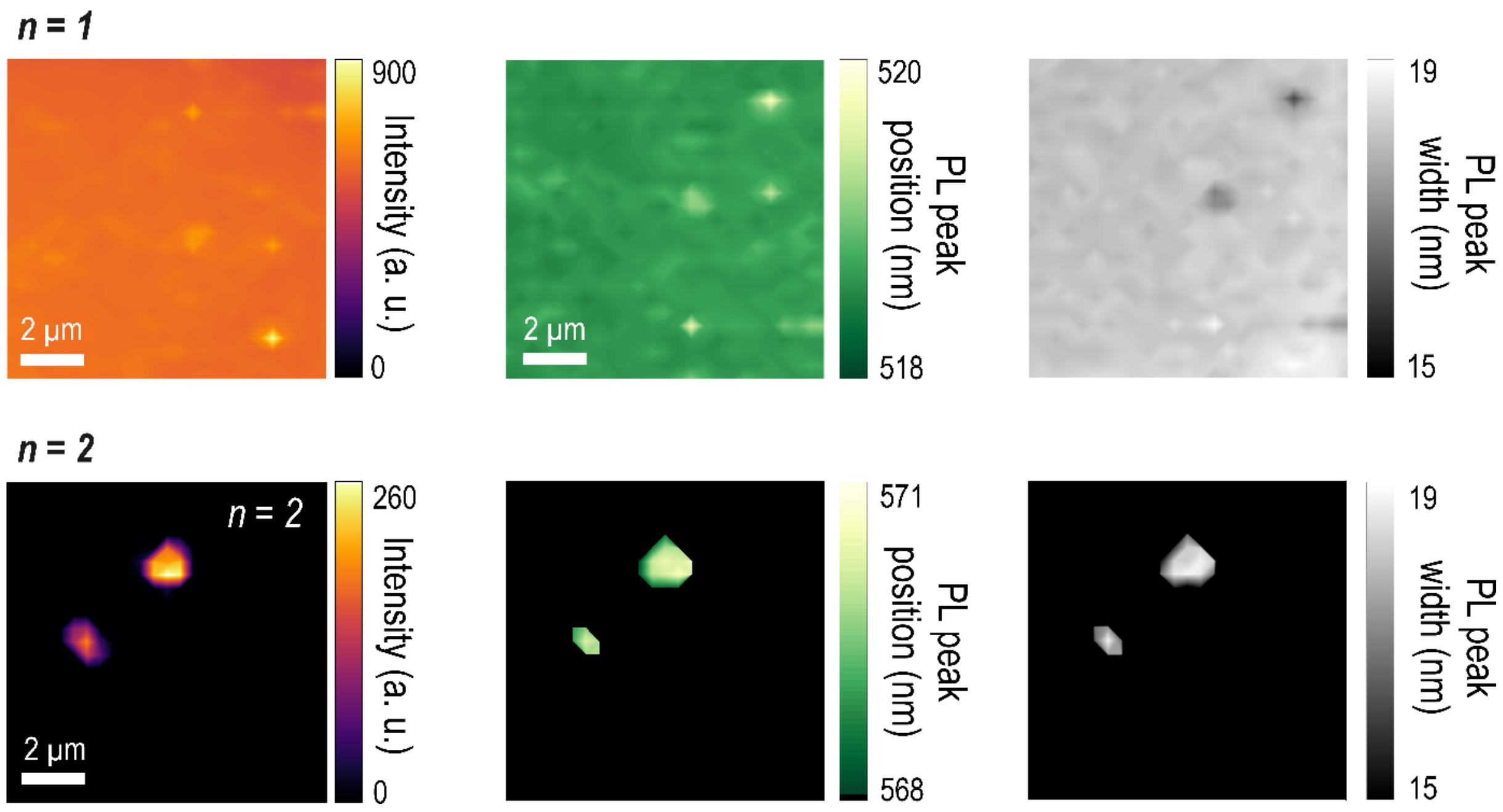

**Figure S5: Results of the photoluminescence fitting of 70 nm thick $(PEA)_2PbI_4$ layer on epi-α-$MAPbI_3$/KCl.** The fitting was with a Gauss function for the $n$ = 1 and $n$ = 2 phases are shown in the upper and lower row, respectively.

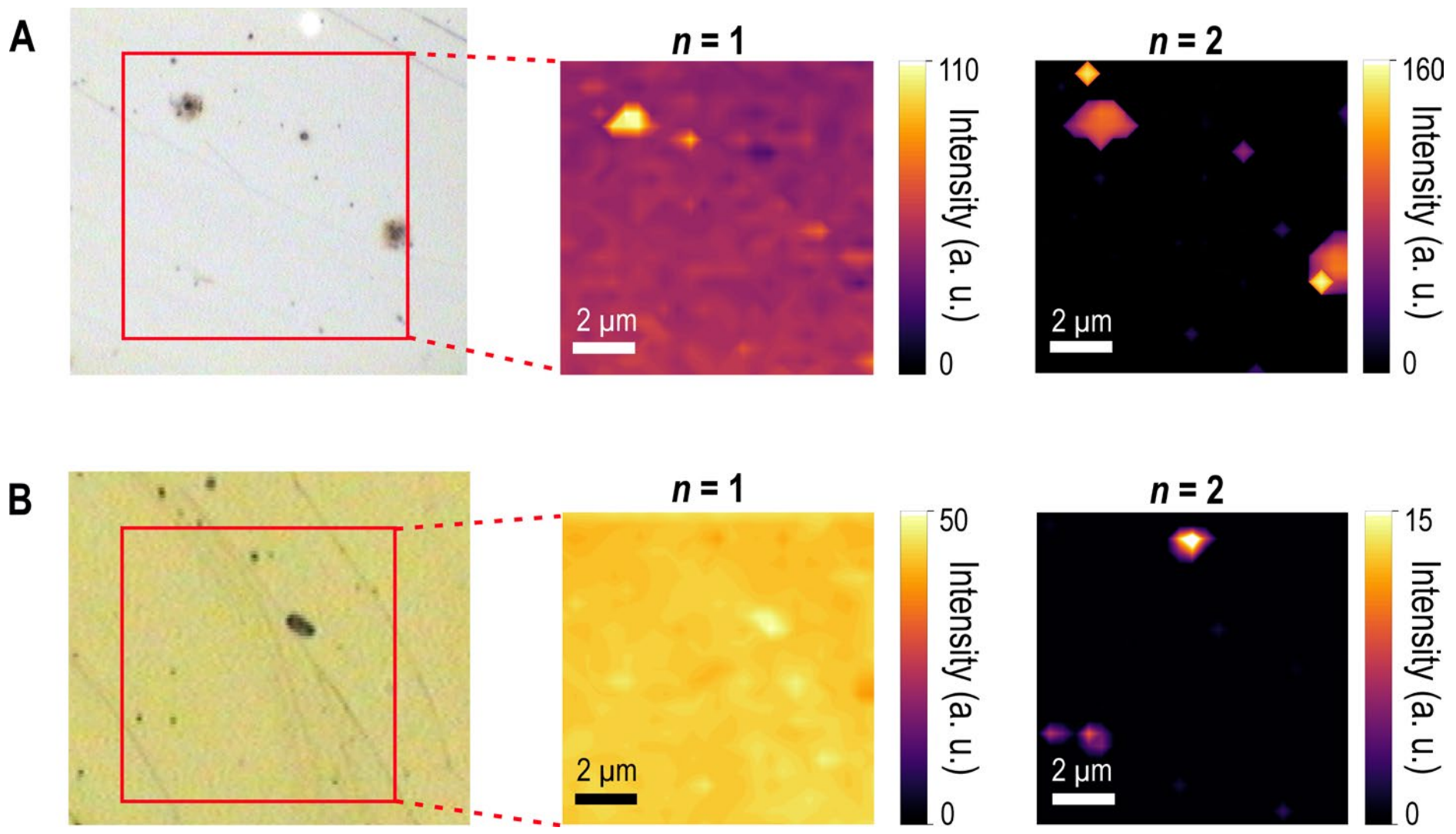

**Figure S6: Spatially resolved PL maps of $(PEA)_2PbI_4$ layers on epi-α-$MAPbI_3$/KCl.** PL mapping was performed in the areas indicated by red squares in the optical microscope image for the $n$ = 1 and $n$ = 2 2D $(PEA)_2PbI_4$ phases at (**A**) 20 nm and (**B**) 40 nm thick $(PEA)_2PbI_4$ layers.

During the PLD process, particulates are known to be generated. [2,3] In Figure S6, these particulates appear under the optical microscope as small, circular dark spots. They typically reside on the film surface and, unlike the surrounding epitaxially strained 3D layer, lack coherent strain. As a result, they tend to act as sites for $n$ = 2 phase formation, which are otherwise absent. It is important to note that the black oval-shaped region visible in Figure S6. B, does not correspond to the one of these particulates, rather it is a surface defect in the underlying KCl substrate.

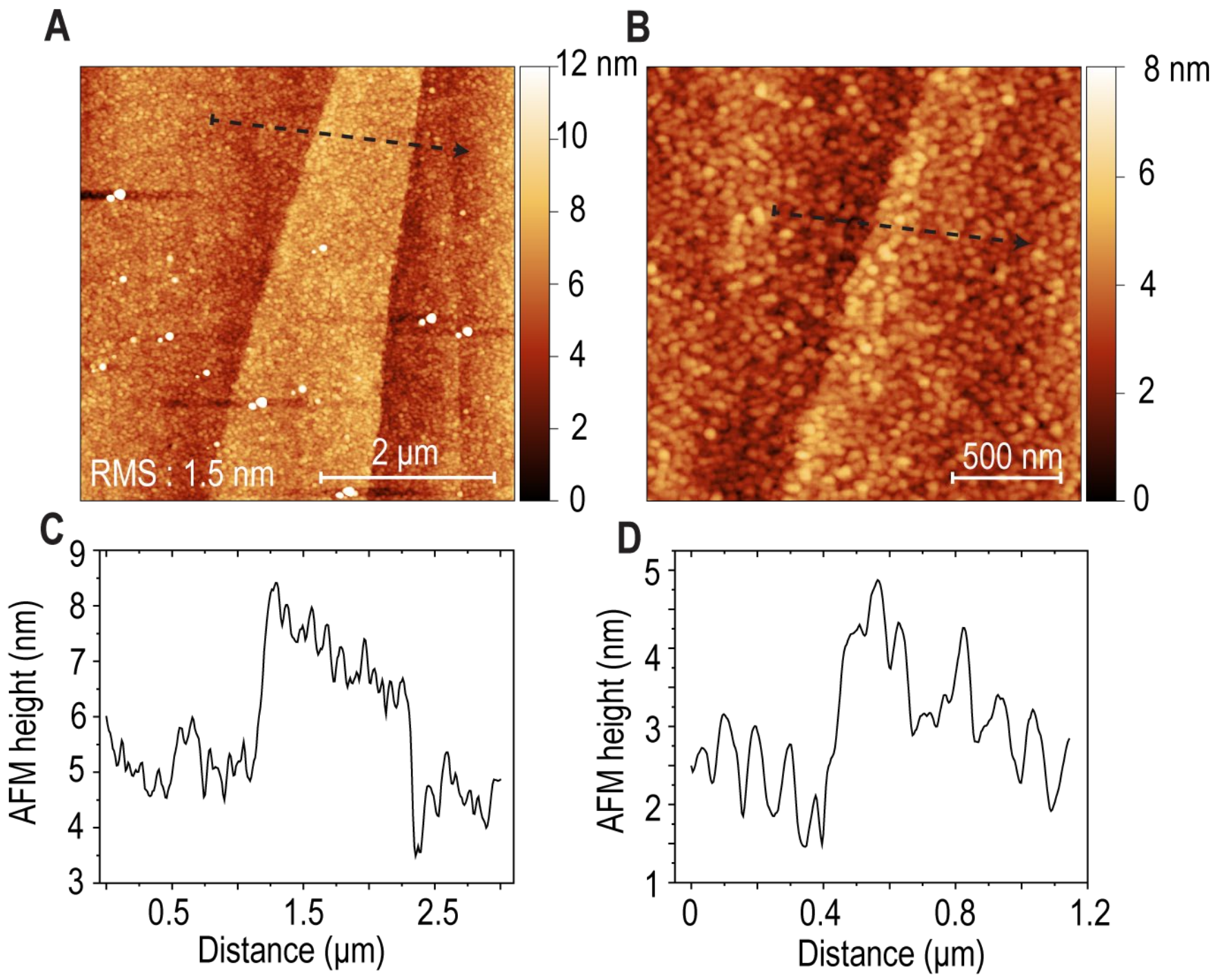

**Figure S7: AFM characterisation of a 20 nm-thick $(PEA)_2PbI_4$ layer on epi-α-$MAPbI_3$ (15 nm)/KCl.** (**A**) 5 x5 $\mu m^2$ and (**B**) a 2 x 2 $\mu m^2$ AFM scans showing the 2D film and the underlying KCl terraces. (**C-D**) Height profiles along the black dashed line in (A) and (B), respectively.

## PLD growth of 2D $(PEA)_2PbI_4$ RP films on different substrates

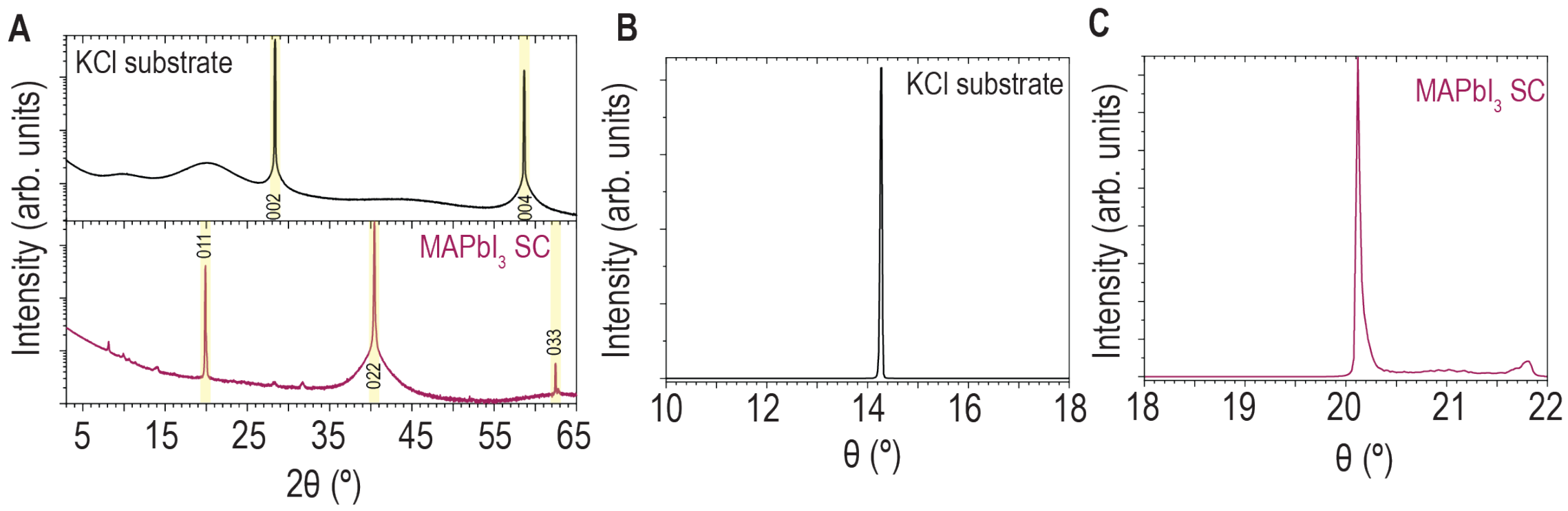

**Figure S8: XRD pattern (A) and rocking measurement for bare KCl (B) and $MAPbI_3$ SC (C) substrate, respectively.**

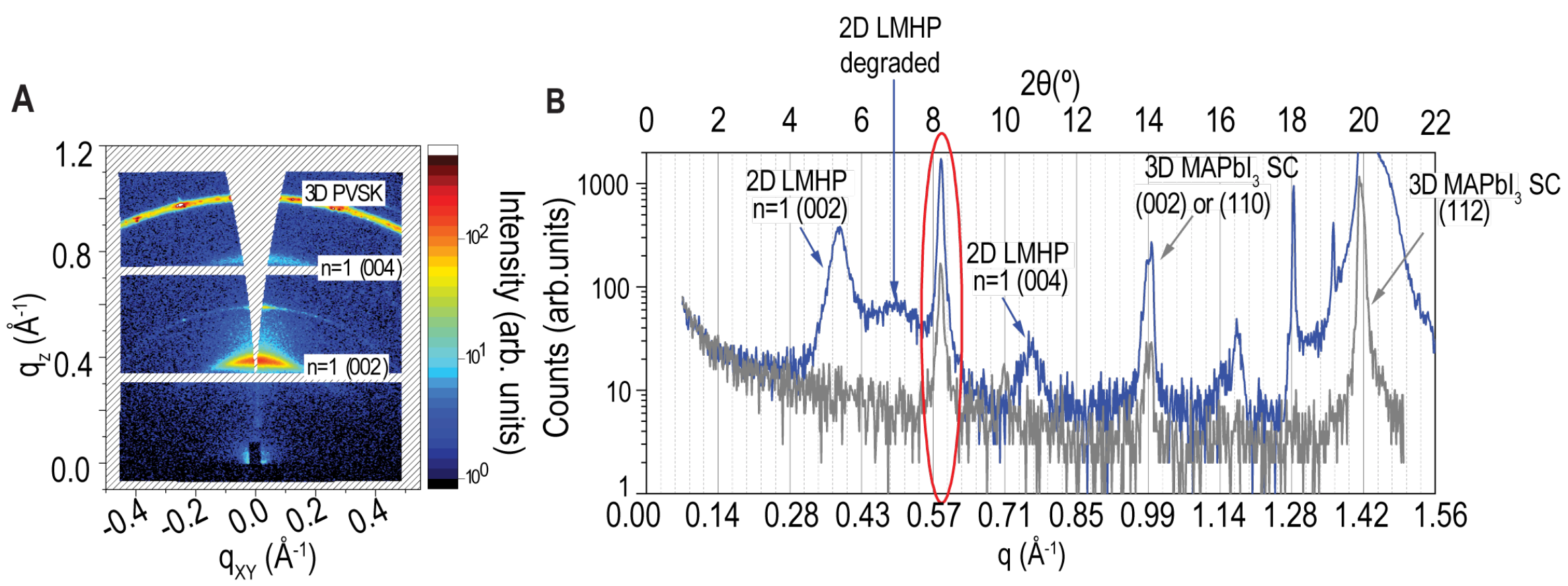

**Figure S9: GIWAXS measurements of 2D $(PEA)_2PbI_4$ grown on top of $MAPbI_3$ single crystal.**

The peak at peak at $q_z \approx 0.6$ Å$^{-1}$ is present at the uncoated side of the $MAPbI_3$ crystallite, indicating that this phase does not come from the 2D perovskite.

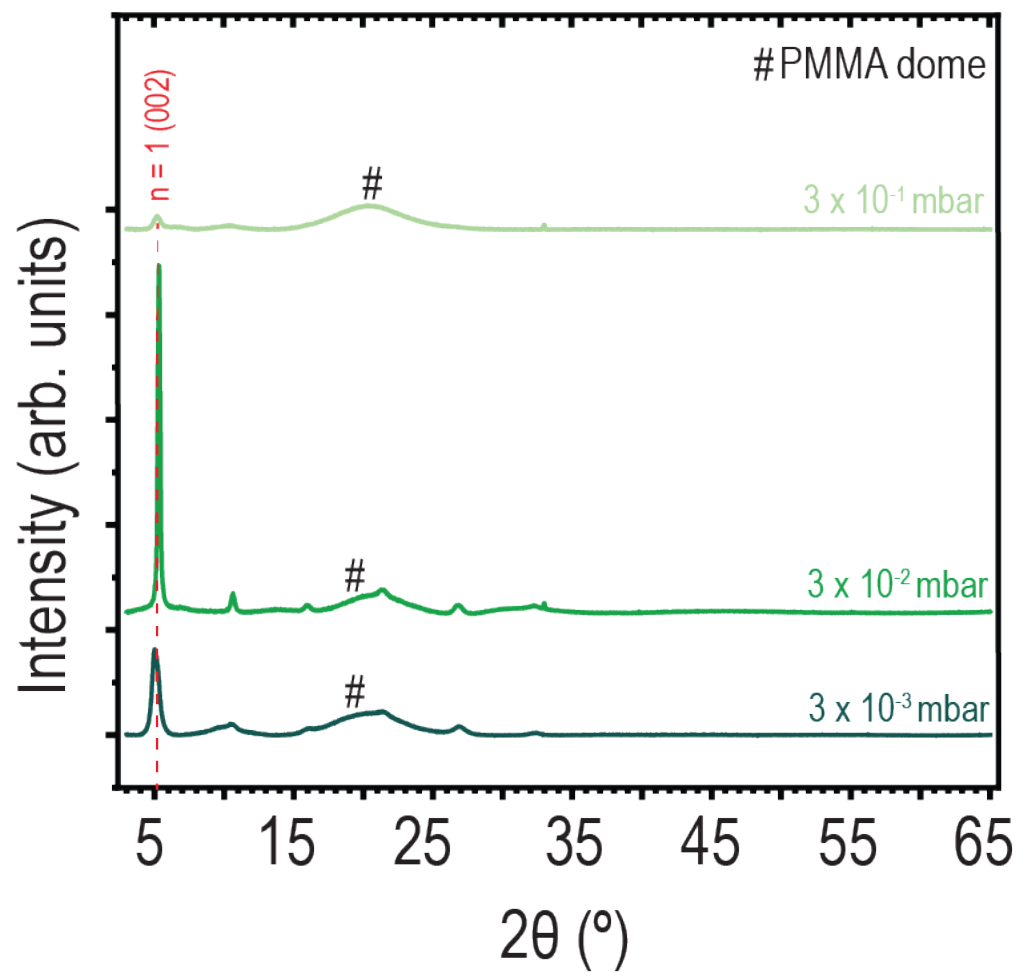

**Figure S10: Different growth pressure to check the correct pressure suited for the growth of the 2D $(PEA)_2PbI_4$ layer.** The substrate used is $SiO_x$/Si.

## Stability monitoring of the 2D $(PEA)_2PbI_4$ layer

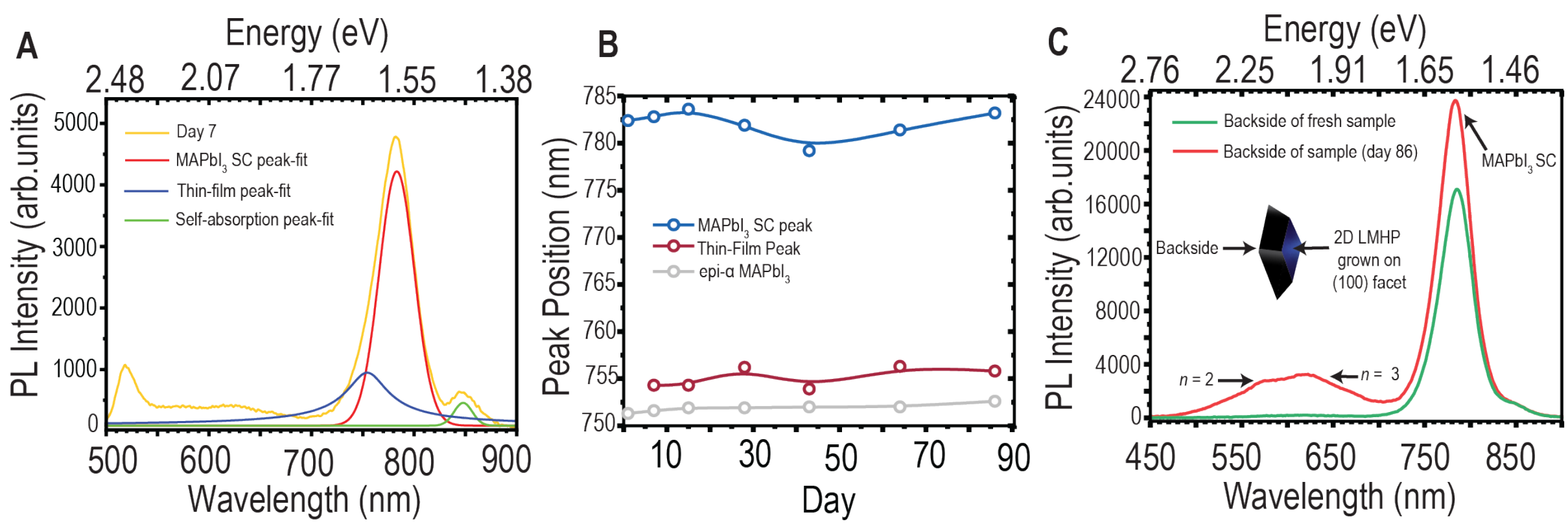

**Figure S11: PL stability analysis** (**A**) Peak-fitting of PL spectra of day 7 of 2D $(PEA)_2PbI_4/MAPbI_3$ (011) SC. (**B**) Peak position vs day for the $MAPbI_3$ SC, thin-film-like layer, and epi-α-$MAPbI_3$. The peak fitting was performed with the Pearson VII function. (**C**) Photoluminescence measured from the backside of a fresh sample with 40 nm 2D layer on $MAPbI_3$ SC and for the sample as shown in the main manuscript after 86 days.

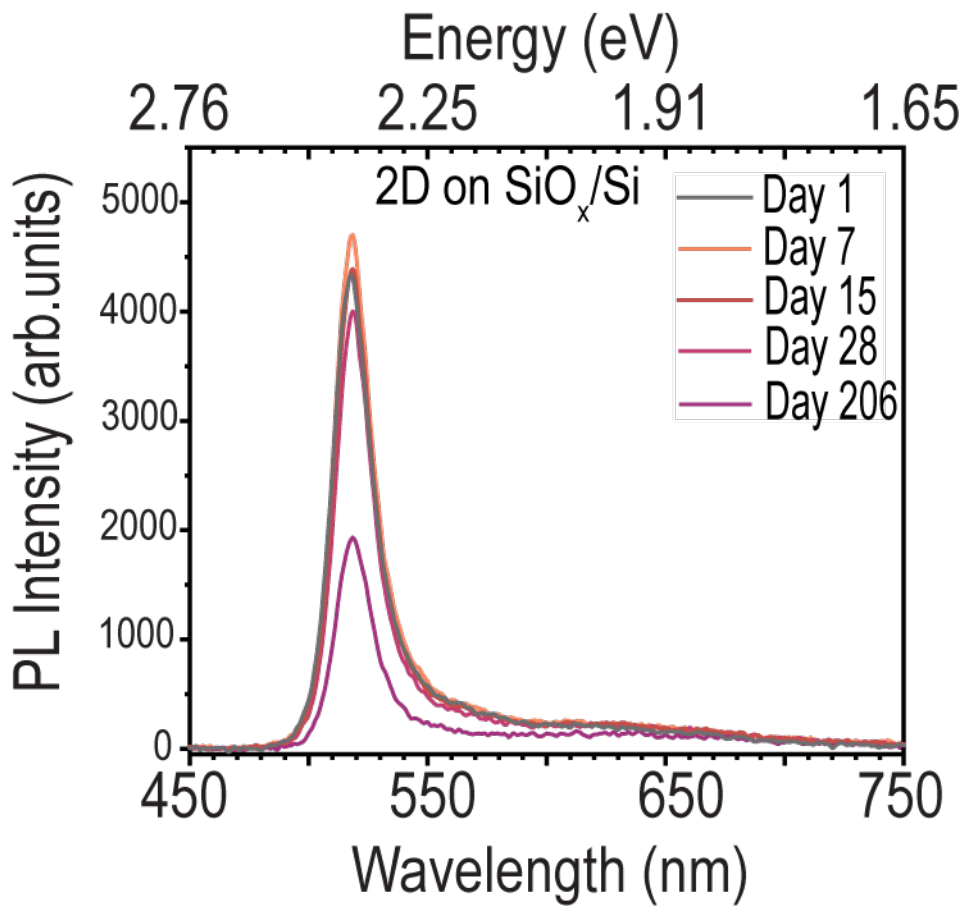

**Figure S12: PL spectra of 40 nm-thick 2D $(PEA)_2PbI_4$ layer grown on $SiO_x$/Si.** The samples were stored and measured inside the glovebox without it being exposed to ambient conditions.